%% file: XMPP1499.tex
\documentclass[conference]{IEEEtran}

\usepackage{ifpdf}
\usepackage[pdftex]{graphicx}
\usepackage{listings}
\begin{document}

\IEEEoverridecommandlockouts

% Title Section

%\title{IEC~61499 Service Interface Function Blocks for XMPP Communication in Smart Grids}
\title{Evaluating XMPP Communication in IEC~61499-based Distributed Energy Applications}

\author{\IEEEauthorblockN{Armin Veichtlbauer, Manuel Parfant, Oliver Langthaler}
\IEEEauthorblockA{Information Technology \& Systems Management\\
Salzburg University of Applied Sciences, Puch/Salzburg, Austria\\
%Email: \{firstname.lastname\}@fh-salzburg.ac.at }
%Email: 
\footnotesize{\{armin.veichtlbauer, mparfant.itsb-m2013, oliver.langthaler\}@fh-salzburg.ac.at }}
\and
\IEEEauthorblockN{Filip Pr\"{o}stl Andr\'{e}n, Thomas Strasser}
\IEEEauthorblockA{Energy Department -- Electric Energy Systems\\
AIT Austrian Institute of Technology, Vienna, Austria\\
%Email: 
\footnotesize{\{filip.proestl-andren, thomas.strasser\}@ait.ac.at }}}

\maketitle

% Abstract Section

\begin{abstract}

The IEC~61499 reference model provides an international standard developed specifically for supporting the creation of distributed event-based automation systems. Functionality is abstracted into function blocks which can be coded graphically as well as via a text-based method. As one of the design goals was the ability to support distributed control applications, communication plays a central role in the IEC~61499 specification. In order to enable the deployment of functionality to distributed platforms, these platforms need to exchange data in a variety of protocols. IEC~61499 realizes the support of these protocols via ``Service Interface Function Blocks'' (SIFBs). In the context of smart grids and energy applications, IEC~61499 could play an important role, as these applications require coordinating several distributed control logics. Yet, the support of grid-related protocols is a pre-condition for a wide-spread utilization of IEC~61499. The eXtensible Messaging and Presence Protocol (XMPP) on the other hand is a well-established protocol for messaging, which has recently been adopted for smart grid communication. Thus, SIFBs for XMPP facilitate distributed control applications, which use XMPP for exchanging all control relevant data, being realized with the help of IEC~61499. This paper introduces the idea of integrating XMPP into SIFBs, demonstrates the prototypical implementation in an open source IEC~61499 platform and provides an evaluation of the feasibility of the result.

\end{abstract}

% no keywords

\IEEEpeerreviewmaketitle

% Chapter Section
% Full Paper: 8 pages IEEE style

\input{Introduction}

\input{Requirements}

\input{Related}

\input{Architecture}

\input{Realization}

\input{Evaluation}

\input{Conclusion}

% Acknowledgment Section

\section*{Acknowledgment}

This work is funded by the Austrian Ministry for Transport, Innovation and Technology (bmvit) and the Austrian Research Promotion Agency (FFG) under the ``ICT of the Future'' program in the OpenNES project (FFG No. 845632).

% Bibliography Section

\bibliographystyle{IEEETran}
\bibliography{AVR,PAM}

\end{document}

%% file: Introduction.tex
\section{Introduction}
\label{sec:introduction}

%This section shall give a short outline of 

%\begin{itemize}
%	\item the motivation of the work at hand: A strong focus here is to be laid on Smart Grid/Home Automation convergence
%	\item the goals to be achieved
%	\item the structure of the rest of the paper
%\end{itemize}

In recent years, the integration of Home Automation (HA) technologies into newly constructed buildings has made some significant progress \cite{Aragues12a}. A remarkable driver for this development is, among others, the emerging smart grid \cite{Tariq12a}. It is based on an underlying Information and Communication Technology (ICT) infrastructure for exchanging control data for distributed sensor/actuator systems \cite{Sood09a}. 

HA, on the other hand, is merely based on local control systems. However, standards in exchanging control data can help setting up HA systems of different vendors \cite{Veichtlbauer12a, Veichtlbauer13a}. Once a standards based HA system is set up, the integration of external data (e.g., weather forecasts, energy prices) to optimize control is the next logical step \cite{Rohjans12a}.

If the owner of a private building additionally intends to generate energy with a Distributed Energy Resource (DER), such as a Photovoltaic (PV) system, he/she is no longer able to separate the structure from the outside world. If an energy surplus is to be injected into the grid, some regulations from the grid operator/energy utility have to be met \cite{SGTF15a}. 

As a consequence, an ICT infrastructure is needed to allow for such regulation. By implementing such an ICT infrastructure, one can also profit from added value, e.g., a reduction of total energy costs enabled by flexible pricing models \cite{Derin10a}.

The drawback of today's solutions still lies in the complexity of realizing distributed control systems. Although there are some promising approaches, available solutions tend to be tailored to concrete applications, instead of being sufficiently generic to allow seamless integration and thus easy interoperability. According to the Smart Grid Architecture Model (SGAM) \cite{M490-12c}, interoperability has to be achieved at different layers for that purpose. 

For the Communication Layer, besides others the eXtensible Messaging and Presence Protocol (XMPP) has been considered a promising protocol \cite{Pichler14} while for the Information Layer, data structures are defined in \cite{Feuerhahn11a,Reinprecht11a}. For the higher layers however, there are few approaches to making functions interoperable.

The most promising idea is an extension of classical Programmable Logic Control (PLC) programming languages, standardized in IEC~61131-3 \cite{IEC61131}, by adding an event-based control mechanism. This is known as the IEC~61499 reference model \cite{IEC61499} and allows for distributed parts of control functionality called ``Function Blocks'' (FBs). 

IEC~61499 uses a special type of such Function Blocks, so called ``Service Interface Function Blocks'' (SIFBs), for integrating functionality used to communicate with the outside world, be it sensors, actuators, or also networking facilities. When using XMPP as underlying protocol, such SIFBs for XMPP have to be used in order to create a standard compliant ICT infrastructure \cite{Parfant15a}.

In this paper, an analysis is performed of the requirements of an ICT infrastructure which is designed to serve as basis for the integration of DER into a smart grid system, while keeping all additional functionality for an optimized HA system (Section~\ref{sec:requirements}). Then, a short overview of technologies which may prove helpful in this context is provided (Section~\ref{sec:related_work}). After describing the concept and the architecture of the proposed solution (Section~\ref{sec:concept}), a short outline of the implemented prototype (Section~\ref{sec:prototype}) is given. Next, the prototype is validated against the requirements and performance tests are conducted (Section~\ref{sec:validation}). Finally, a brief conclusion and an outlook of the planned further research work is given (Section~\ref{sec:conclusions}).

%% file: Requirements.tex
\section{Communication Requirements in Smart Grids}
\label{sec:requirements}

% In this section we have to clearly point out why this work is especially relevant for the Energy Sector.

In the areas of smart grids and HA, a plethora of different use cases can be found. According to \cite{Aragues12a}, they can roughly be divided into Ambient Assisted Living (AAL), Home Automation, and Energy Management. Some use cases can be assigned to several of these categories. 

AAL uses cases contain, besides others, detection of emergency situations, automatic transfer of health data, or remote unlocking of doors \cite{VDE08a}. Typical use cases in HA are Heating, Ventilation, and Air Conditioning (HVAC) control, but also alarms in case of fires or intrusions \cite{Rovsing11a}. The automatic shedding of loads as well as reminder functions for active loads share aspects with the last use case category, Energy Management. Further examples in this category are advertisements of the energy rate or the current energy consumption, load management, and demand side management \cite{CEN12a}.

From the previously defined use cases, the requirements for the intended ICT infrastructure can be derived. Similar processes can be found in \cite{Parra09a} or \cite{Kamilaris11a}. The IntelliGrid method, originally developed by the Electric Power Research Institute (EPRI), also receives a great deal of interest in the power and energy domain for the specification and analysis of use cases \cite{IEC62559}. In the given case, the focus has been on the Energy Management use cases, as these are of special interest for the setup of a smart grid system; yet other use cases may additionally be covered by the resulting requirements. As the ICT infrastructure consists of two complementing categories, i.e., the engineering environment for distributed grid applications and services and the underlying communication network and corresponding protocol stack, the requirements have been divided accordingly.

For the engineering process the following main requirements can be derived \cite{Parfant15a}:

\begin{itemize}
	\item{\textit{[REQ01]} The engineering environment must support the development of distributed applications.}
	\item{\textit{[REQ02]} The development process must be as simple as possible.}
	\item{\textit{[REQ03]} The execution system must offer sufficient performance for the execution of distributed applications.}
	\item{\textit{[REQ04]} The integration and connection of legacy systems/components must be possible.}
	\item{\textit{[REQ05]} The support of default interfaces (e.g., TCP/IP, XMPP) must be possible.}
	\item{\textit{[REQ06]} The engineering environment has to support platform independent development.}
\end{itemize}

From the communication point of view, the main requirements are \cite{Parfant15a}:

\begin{itemize}
	\item{\textit{[REQ07]} The connection of new nodes must be as simple as possible (e.g., Plug \& Play).}
	\item{\textit{[REQ08]} The protocol must not restrict the variability of the payload.}
	\item{\textit{[REQ09]} The protocol must support asynchronous communication (publish-subscribe and push).}
	\item{\textit{[REQ10]} The protocol must support synchronous communication (request-response and client-server).}
	\item{\textit{[REQ11]} The protocol must offer strong encryption.}
	\item{\textit{[REQ12]} The protocol must offer authentication.}
 	\item{\textit{[REQ13]} The protocol must be able to address local and external nodes.}
	\item{\textit{[REQ14]} The protocol must offer sufficient performance for the execution of the applications.}
	\item{\textit{[REQ15]} The protocol must support data transfer between heterogeneous systems.}
	\item{\textit{[REQ16]} Web services and service-oriented architectures must be supported to facilitate connecting to the Internet.}
\end{itemize}

%% file: Related.tex
\section{Related Work}
\label{sec:related_work}

%This section shall describe the scientific state of the art and commercial tooling in the relevant fields for the work at hand, as well as their usefulness to the defined requirements:

%\begin{itemize}
%	\item Which automation standards have been considered, why is 61499 suitable to distributed automation (and thus for Smart Grids) \cite{Vyatkin11a}, which other solutions 
%	in scientific or commercial work have been researched 
%	\item Which high layer protocols have been considered, why is XMPP suitable to middleware frameworks for Smart Grids, also a bit of SOA and ROA discussion should 
%	be put here \cite{Pichler15a}, which other solutions in scientific or commercial work have been researched 
%	\item Which coding schemes have been considered, why is a combination of ASN.1 with Base64 Encoding used for the prototype, what has been done in comparable environments,
%	e.g. mail, etc.
%\end{itemize}

%[ insert Sota from AIT (ca. 1/2 to 1 page): 
%
%\begin{itemize}
	%\item 61131
	%\item 61499 (e.g. \cite{Vyatkin11a})
	%\item XMPP vs. alternatives
	%\item SOA vs. ROA (e.g., \cite{Melik12a})
	%\item etc.
%\end{itemize}
%
%Hereby it is important to highlight, which considered technologies meet our requirements best! ]

\subsection{Distributed Automation according to IEC~61499}

In the automation domain, the most commonly used control approach is IEC~61131-3, which has been developed to provide a common model for the realization of PLCs \cite{IEC61131}. It mainly targets control systems consisting of one or more tightly coupled controllers, meaning that data shared between the devices is replicated by an underlying service.  

In contrast, IEC~61499 provides an automation approach which has been developed to model distributed industrial process measurement and control systems with a main focus on multi-vendor support. It defines a reference architecture for distributed automation and control systems used in industrial environments (manufacturing systems, power and energy systems, building automation, etc.). Differing from other automation standards like IEC~61131, this approach focuses on the entire control solution; it covers the modeling of control applications, provides abstract interfaces for communication and process interaction, but also briefly describes the underlying distributed hardware setup \cite{IEC61499,Vyatkin11a}.

\begin{figure}[!htbp]
	\centering
		\includegraphics[width=0.98\columnwidth]{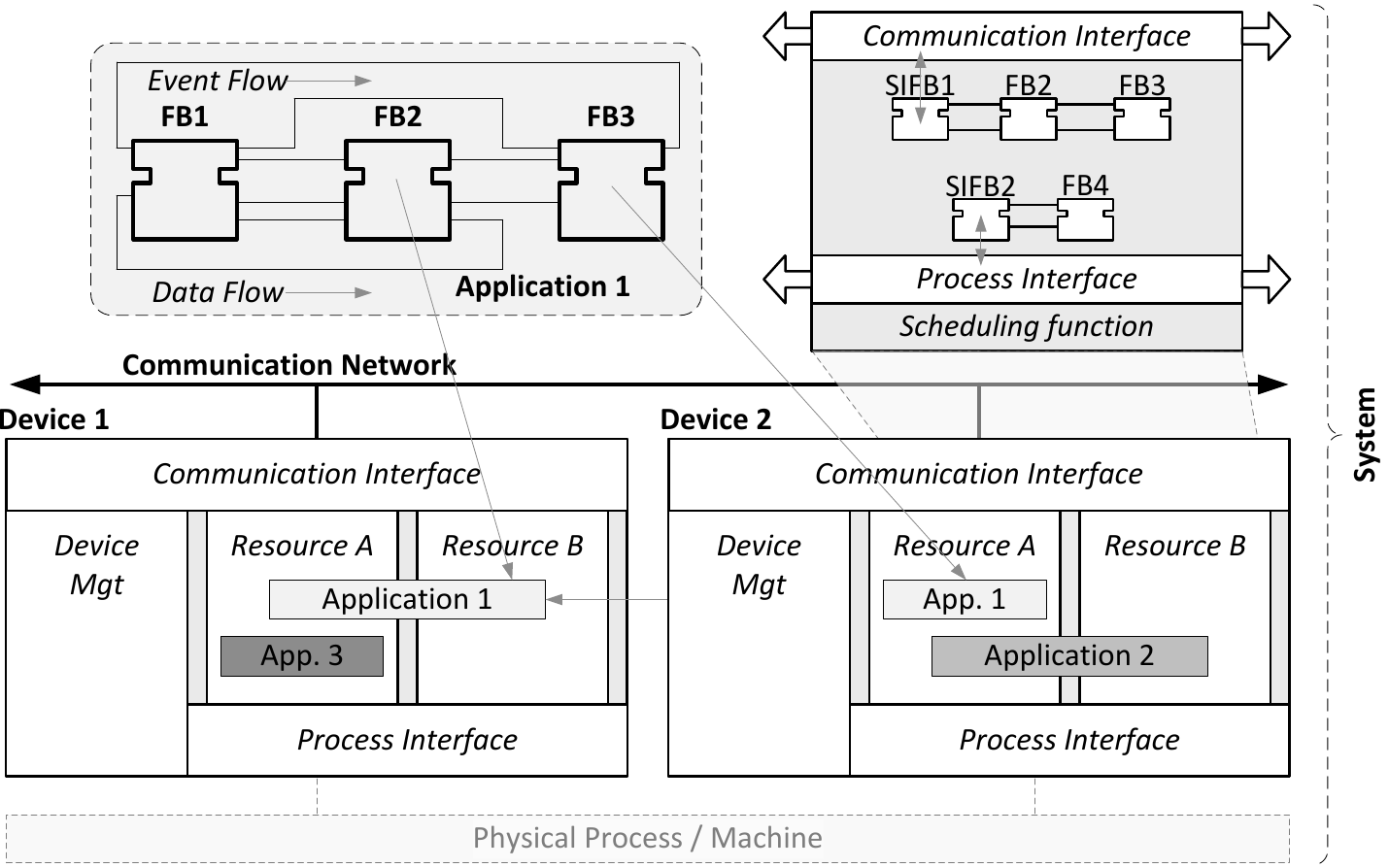}
	\caption{IEC~61499 reference model for distributed automation \cite{IEC61499,Vyatkin11a}.}
	\label{fig:iec61499}
\end{figure}

The core modeling elements are Function Blocks (FB), which encapsulate modular control software into components. The IEC~61499 FB model is based on its predecessor IEC~61131-3, but uses events for defining the execution flow between the FBs. Multiple FBs connected to an FB network make up an IEC~61499 control application. Once an application has been defined, it can be deployed to intelligent field devices, referred to as Devices in the standard \cite{IEC61499}. Access to the communication network as well as to the underlying controlled process is encapsulated into special FBs called SIFBs. A brief overview of the IEC~61499 concept with the System, Application, Device, Resource, Communication Network, and FB (incl. SIFB) models is provided in Fig.~\ref{fig:iec61499}.

The integration of different communication protocols into IEC~61499 communication SIFBs has already been demonstrated by several research groups \cite{Andren:2011,Christensen:2011URL,Scarlett:2011,Schwab:2004,Weehuizen:2007}. All of them have in common that they primarily address industrial automation applications. Smart grid related applications have so far mainly been targeted using the IEC~61850 interoperability approach \cite{Andren:2014,Higgins2011}.  

%\subsection{Software Architectures}
%
%Service-oriented Architecture (SOA) is an approach from computer science. It is based on providing and requesting services. A service is usually a software entity which encapsulates business or control logic as well as functionalities provides by a component (which can be internally represented by an agent). SOA-based approaches are not very common in the power and energy systems domain today. However, a good and more comprehensive overview of using SOA-based approaches in the energy domain is provided by Vrba et al. \cite{Vrba14}. Furthermore, the IEC suggests a Seamless Integration Reference Architecture (SIA) based on SOA principles used in the domain of smart grids. 
%
%\textbf{ToDo: SOA vs. ROA (e.g., \cite{Melik12a})}

\subsection{Communication Approaches}

Modern communication approaches such as service-oriented arcitectures and web services are common in Internet and cloud applications. They now receive increasing attention for the realization of smart grid systems \cite{Warmer09}. In this domain, the SOAP based ``Devices Profile for Web Services'' (DPWS) as well as the Representational State Transfer (REST) provide important approaches when implemented in embedded controllers \cite{AESOP14}. DPWS mainly deals with the ubiquitous device integration using web services embedded in distributed devices, whereas REST can be seen as an alternative integration approach which emerged from the world wide web initiative and focuses mainly on the M2M communication.

Furthermore, a very interesting approach which receives increasing attention in the power and energy domain is XMPP. It is a message oriented middleware protocol located on top of existing TCP connections, secured with Transport Layer Security (TLS) end-to-end encryption. It can thus be seen as a layer 5/6 protocol with respect to the ISO's Open Systems Interconnection (OSI) reference model \cite{OSI7498-94a}. The XMPP core definition is specified in RFC6120 \cite{Saint11}. Its aim is to distribute structured data (XML based ``XMPP stanzas'') via a network between two or more entities. Hereby, a fundamental distinction is made between server and client entities. Clients have to register at a server instance and can then send data to any XMPP server which is accessible via TCP, or to clients which are registered at one of these servers. This procedure is very similar to the email system. Compared to web services, XMPP has the advantage of allowing for publish/subscribe communication patterns, a functionality not natively supported by web services and mostly emulated by polling in regular intervals. XMPP on the other hand uses ports which are often blocked, whereas Web services most often rely on http or https and thus only require ports 80, 8080 or 443 to be open.

\subsection{Comparison with Communication Requirements}

Regarding the application engineering part, IEC~61499 is perfectly fitting the corresponding requirements REQ01--06. Although IEC~61131-3-based solutions are in principle possible, IEC~61499 has the advantage of being natively developed for distributed systems, whereas the tight coupling of IEC~61131-3 is in contrast to common service-oriented approaches in the smart grid domain and can thus fulfill REQ01 only partly.

As for the communication part, XMPP natively supports asynchronous (publish/subscribe) communication in contrast to web services, which need additional enhancements like DPWS. Also, authentication is integrated natively. As drawback, XMPP is a protocol with a rather big footprint; for lightweight implementations this could lead to some performance problems. The most important point with XMPP is yet the support by big industry players in the smart grid domain, as well as in standardization bodies like IEC. In IEC~TC57~WG21, XMPP is foreseen as transport protocol between ``Customer Energy Management'' (CEM) systems and external entities \cite{Pichler14}. 

Summarizing, a combination of IEC~61499 (as a means for the engineering of smart grid applications) and XMPP (as a means for the communication of distributed parts of the grid) provides all required functionality and appears thus to be a feasible approach for a standardized and interoperable ICT infrastructure.

%% file: Architecture.tex
\section{Prototype Architecture and Test Concept}
\label{sec:concept}

\subsection{Test Cases and Architecture}

In order to verify the fulfillment of the aforementioned requirements, two test cases have been defined:

\begin{itemize}
	\item{\textit{[TC1] Simple Demand Side Management function with local storage:} An energy storage system (i.e., a buffer battery) is charged and discharged according to the current voltage values (over-voltage, normal voltage, and under-voltage) of the electricity grid. This test case uses an asynchronous connection between the network operator (NetOp) and a CEM controlling the buffer battery, i.e., the NetOp publishes voltage conditions on change, and the CEM subscribes to these condition changes.}
	\item{\textit{[TC2] Reminder function:} The states of electrical loads in the household are shown on an appropriate display. This test case uses a synchronous connection between a CEM controlling the electrical load and the display, i.e., the display requests the current state of the loads in a fixed time interval, and the CEM responds to this.}
\end{itemize}

%\subsection{Architectural Considerations}

Subsequently, a simple prototypical architecture covering these test cases has been defined. The architecture consists of three main system components:

\begin{itemize}
	\item \textit{The Network Operator (NetOp):} This part is a very simple emulation of a network operator. Via three push buttons, the voltage conditions (over-voltage, normal voltage, and under-voltage) can be triggered, which are then sent to the CEM.
	\item \textit{The CEM:} This part of the setup emulates an energy storage system (i.e., a buffer battery), which charges/discharges depending on the current voltage level of the \mbox{NetOp}, and an electrical consumer, which can be triggered via a push button.
	\item \textit{The Display:} This part visualizes the presence of an electrical load at the CEM via a Light Emitting Diode (LED).
\end{itemize}

All of these components communicate via XMPP. For reasons of performance and for having a central instance for rights management, an additional XMPP server had to be set up. This should not be confused with an application's notion of client and server; as for XMPP, in many cases the two logical endpoints of a communication activity are both implemented as XMPP clients, with central XMPP servers for switching stanzas from sender to receiver. The IEC~61499 functionality is thus only needed in the endpoints, i.e., the three main components. The XMPP server simply forwards XMPP stanzas and does not need any knowledge about the IEC~61499-based stanza payload.

\begin{figure}[!htbp]
	\centering
		\includegraphics[width=0.98\columnwidth]{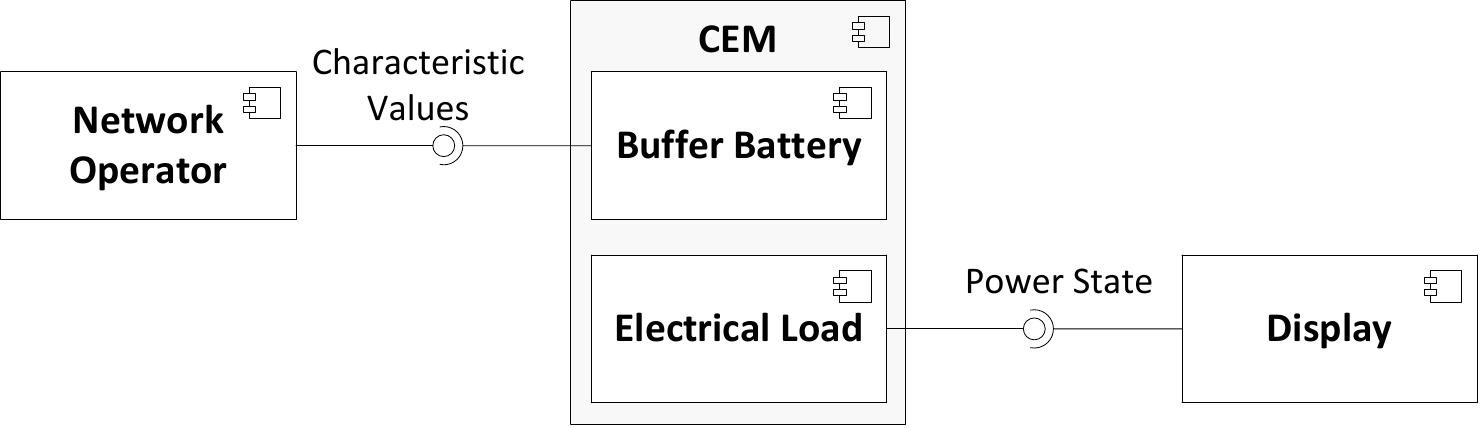}
	\caption{Component diagram representation of the prototype architecture.}
	\label{fig:componentdiagram}
\end{figure}

In order to be able to execute the test cases, the functionality had to be mapped onto the components of the prototype architecture, as shown in the component diagram in Fig.~\ref{fig:componentdiagram}.

% The test case \textit{TC1} is executed on the NetOp and the CEM. The NetOp detects the characteristic values (over-voltage, normal voltage, and under-voltage) of the grid and sends them to the CEM. The CEM charges or discharges the buffer battery according to the transmitted values. The second test case \textit{TC2} is executed on the CEM and the display. The display cyclically requests the power state of the electrical load from the CEM and visualizes the state.

\subsection{Validation Environment}

For testing the proposed XMPP integration into IEC~61499 SIFBs, an appropriate validation environment is needed, i.e., the three main components have to be realized with running XMPP clients and an IEC~61499 execution system on it. Such a standard-compliant runtime environment has to provide corresponding communication SIFBs which are able to use the respective XMPP clients for publish/subscribe and for client/server (request/response) communication. 

%Raspberry Pi are used for the implementation of these systems because this embedded device is cheap and small and offers sufficient performance. The electricity grid, the storage battery and the display were simulated by test boards with keys and LEDs, see Figure \ref{fig:testenv}. 
%The test board of the network operator consists of three buttons and three LEDs. The characteristic value can be set by the three buttons and is illustrated by the three LEDs (red - overvoltage, green - normalvoltage, yellow - undervoltage). The test board of the CEM consists of three LEDs and one button. The upper two LEDs illustrates the charging (red) and discharging state (green) of the buffer battery and the third LED (yellow) the power state of the electrical load. The electrical load can be switched on and off with the help of the button. The test board of the display consists of one LED (yellow). This LED illustrated the power state of the electrical load.

Thereto, embedded control devices have been used. Raspberry Pis have been selected because they are inexpensive, compact, and offer sufficient performance. For the Raspberry Pi platform, there are several Operating Systems (OSs) available: various Linux distributions and one RISC OS version. 
%The most widespread and best supported OS for the Raspberry Pi is Raspbian, which is based on Debian. This OS was used for the test environment due to its maturity.
The Debian based Raspbian was used for the test environment due to its maturity.

For the IEC~61499 development and runtime environment, the range of available and currently maintained tools is limited. The commercial tools ISaGRAF Workbench and nxtStudio as well as the open source tools FBDK and 4DIAC are available \cite{Christensen12a}. 4DIAC was chosen because it offers a large range of functions, FBs for the General Purpose Input/Output (GPIO) pins on the Raspberry Pi embedded controller, and a layer model for communication \cite{Zoitl:2010}.

%The final two components are the XMPP client library and the XMPP server. There are many client libraries and servers available; a list can be found on the website of the XMPP Standards Foundation (XSF)\footnote{http://xmpp.org/software}. 

For the XMPP client library, important factors are the range of functions, the quality of the Application Programming Interface (API) documentation, a small footprint and frequent maintenance releases. Besides, it is desirable for the library to be developed in C/C++, the same programming language as the 4DIAC runtime system FORTE. These requirements are best met by the gloox library, which was thus used for the development. The requirements for the XMPP server are almost identical. It must have a small footprint, frequent maintenance releases, a good documentation and must be easy to configure. Ejabberd was chosen because it has a web interface for the configuration, it is available as a package on Raspbian, and it is well established.

\begin{figure}[!htbp]
	\centering
		\includegraphics[width=0.68\columnwidth]{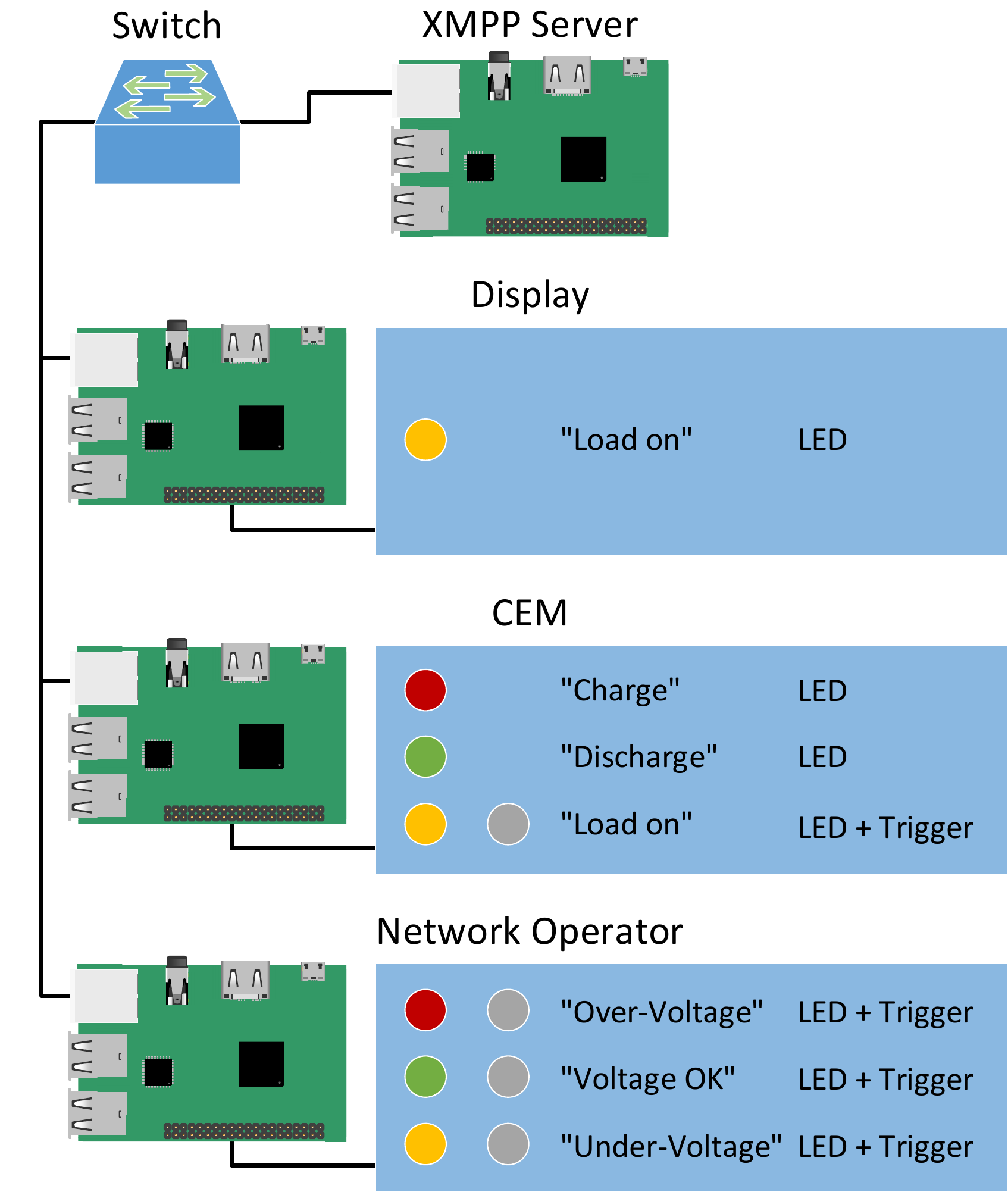}
	\caption{Test environment using Raspberry Pi embedded controllers.}
	\label{fig:testenv}
\end{figure}

Summarizing, for the validation of the test cases \textit{TC1} and \textit{TC2}, a test setup as shown in Fig.~\ref{fig:testenv} has been used. Raspberry Pi controllers are used in combination with extension boards emulating the behavior of NetOp, CEM, and display, using LEDs and push buttons. The test boards for the simulation of the electricity grid, buffer battery and electrical load are connected to the Raspberry Pis via their GPIO pins.

%% file: Realization.tex
\section{Prototypical Realization}
\label{sec:prototype}

% This section shall give a short overview of the implemented prototype and the programming environment.

\subsection{IEC~61499/4DIAC Communication Infrastructure}

The existing communication functionality of FORTE comprises a layered model. Thus, it is for instance possible that the first layer encodes the data (e.g., Abstract Syntax Notation One (ASN.1)), the second layer encrypts the encoded data (e.g., Advanced Encryption Standard (AES)), and the third layer transmits the encrypted data (e.g., TCP/IP). Hereby, FORTE supports the generic communication types ``publish/subscribe'' (asynchronous) and ``client/server'' (synchronous). As transport protocol, publish/subscribe uses UDP, whereas client/server uses TCP.

\begin{figure}[!htbp]
	\centering
		\includegraphics[width=0.98\columnwidth]{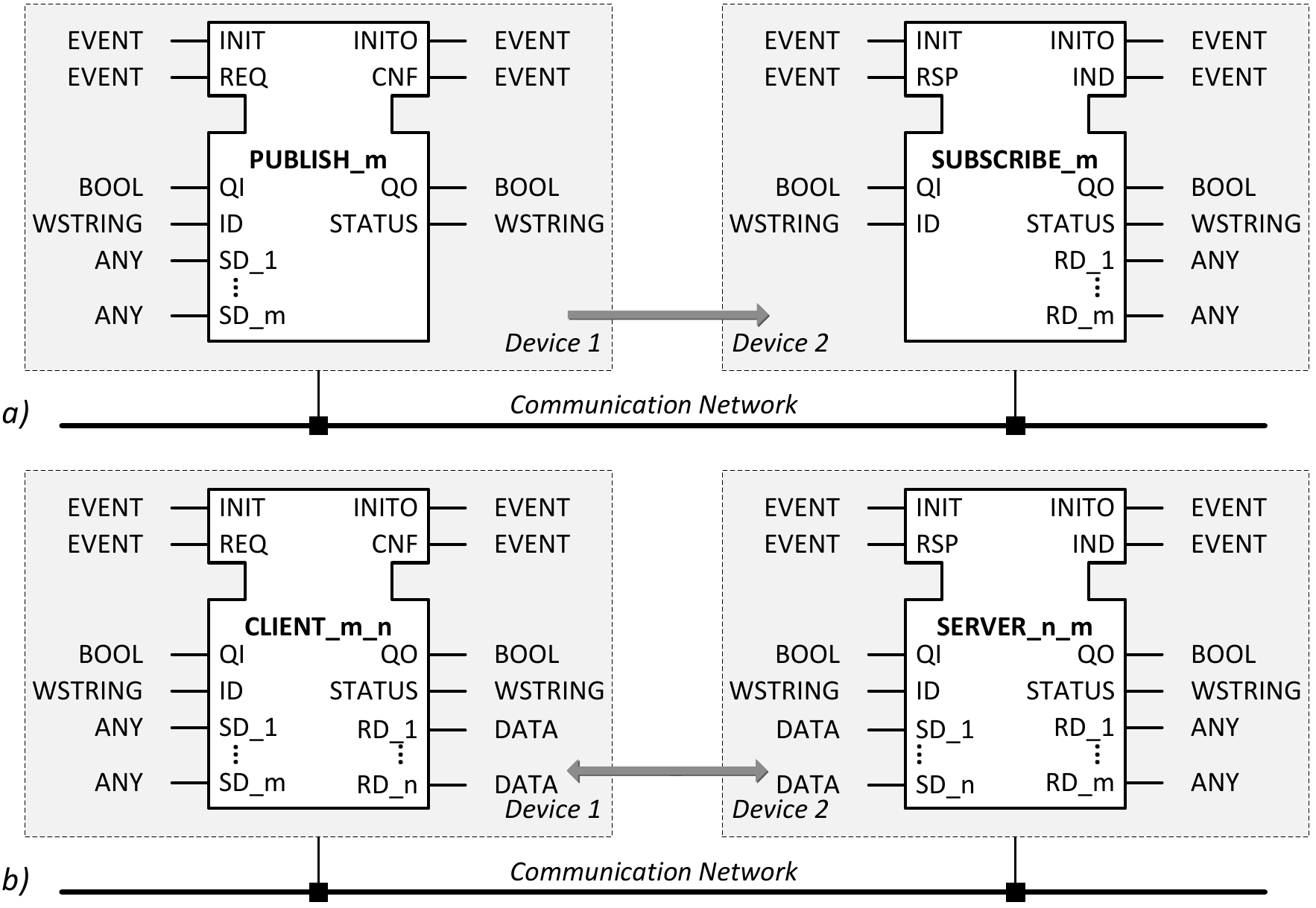}
	\caption{IEC~61499 communication SIFB patterns \cite{IEC61499}: a) ``publish/subscribe'' (asynchronous), b) ``client/server'' (synchronous).}
	\label{fig:iec61499_csifb}
\end{figure}

These communication functionalities can be utilized via the pre-defined communication FBs \texttt{publish} and \texttt{subscribe}, respectively \texttt{client} and \texttt{server}, as shown in Fig.~\ref{fig:iec61499_csifb}. These SIFBs are configured via the \texttt{ID} parameter, which consists of one or more protocols with optional parameters (e.g., \texttt{fbdk[].ip[192.168.20.1:61499]}). In this example, \texttt{fbdk} defines ASN.1 as the protocol of the first layer, and \texttt{ip} defines TCP/IP or UDP/IP with the parameters \texttt{IP address} and \texttt{port} as the protocol (stack) for the second layer.

\subsection{XMPP Communication Integration}

FORTE's communication layer model can be utilized for the implementation of XMPP communication. The existing ASN.1 protocol can be used for data encoding, i.e., FORTE uses ASN.1's Basic Encoding Rules (BER) for serialization of structured data. For the actual XMPP data transfer, a new communication layer had to be developed. This solution permits highest flexibility from the engineering point of view, as developers are enabled to change the protocol for the data transfer or to add an additional encryption layer simply by changing the \texttt{ID} input of the communication FB. 

The output of the BER serialization is a byte stream, whereas XMPP is a messaging protocol only transmitting Unicode characters. For that reason, each byte stream has to be converted into a character stream. The gloox library supports Base64 encoding \cite{RFC4648-06a}, which is therefore used for the conversion. Alternatively, ASN.1's XML Encoding Rules (XER) could have been used for directly producing XML code which could be transported as XMPP stanza; yet this encoding rule is not currently supported by the FORTE runtime.

%\textbf{ToDo: Is XMPP used for encapsulating the ASN.1 message as data payload? The relationship between ASN.1 and XMPP should be better explained.}

\vspace{2mm}

\subsubsection{Publish/Subscribe}

The publish/subscribe communication type uses XMPP ``Presence Stanzas'' for data transfer. When a connection is established, the subscriber sends a subscription request to the publisher. The publisher accepts the request and adds the subscriber to its contacts list (roster). The publisher sends messages to all contacts of its roster.

The format of the \texttt{ID} input string for the \texttt{publish} FB is 

\begin{lstlisting}
fbdk[].xmpp[encryption:publisher full JID:password:XMPP server IP address]
\end{lstlisting}

and for the \texttt{subscribe} FB 

\begin{lstlisting}
fbdk[].xmpp[encryption:subscriber full JID:password:XMPP server IP address:publisher full JID]
\end{lstlisting}

The parameter \texttt{encryption} defines the XMPP encryption (0 -- none, 1 -- TLS). In the concrete setup used for evaluation, the ID input string for the \texttt{subscribe} SIFB was defined as follows (localhost is the XMPP server here with the IP address 192.168.1.210):

\begin{lstlisting}
fbdk[].xmpp[1:cemdsm@localhost/res:***: 192.168.1.210:netop@localhost/res]
\end{lstlisting}

%\textbf{ToDo: A generic example is needed with SIFBs and their IDs; use the below listing and fill in the correct configuration string for XMPP}

\subsubsection{Client/Server}

The client/server communication type uses XMPP ``IQ Stanzas'' for data transfer. Alternatively, ``Message Stanzas'' could be used; yet these have the disadvantage (compared to IQ Stanzas) that no reply mechanism is enforced for this stanza type. Roster management is not necessary for server/client communication, thus messages can be sent without a valid subscription.

The format of the \texttt{ID} input string for the \texttt{client} FB is 

\begin{lstlisting}
xmpp[encryption:client full JID:password:XMPP server IP address:server full JID]
\end{lstlisting}

and for the \texttt{server} FB

\begin{lstlisting}
xmpp[encryption:server full JID:password:XMPP server IP address:client full JID]
\end{lstlisting}

%\textbf{ToDo:It should be explained that an external XMPP server is needed and that the SERVER FB is implemented as an XMPP client; It is not clear how to use the CLIENT/SERVER pattern of IEC~61499, when should the CLIENT FB be used and when should the SERVER FB be used; CLIENT/CLIENT connections do not make sense in IEC~61499, how should an XMPP client to XMPP client connection be represented in IEC~61499;}

An own XML namespace named \texttt{forte} is used for data transmission. This namespace contains an XML element \texttt{Value} for the transmission of base64 encoded data.

Summarizing, the publish/subscribe pattern should be applied for pushing data from one device to another one, whereas client/server is mainly used for such kind of use cases where a device provides services to another one.

%Finally, the following listing provides examples for configuring the \texttt{publish}/\texttt{subscribe} SIFBs which have been used for the realization of the prototype architecture in the 4DIAC framework as explained above.

%% file: Evaluation.tex
\section{Validation and Evaluation Results}
\label{sec:validation}

% This section shall give a short summary of validation tests as defined with the test cases of the requirements chapter and compare the results with the respective requirements.
% Furthermore all additional quality tests (e.g., scalability, performance) shall be mentioned here.

%This section covers the discussion of the implemented test applications as well as the functional validation and performance evaluation of test cases \textit{TC1} (asynchronous communication with publish/subscribe) and \textit{TC2} (synchronous communication with client/server).

\subsection{Test Applications}

For the execution of the two test cases \textit{TC1} (asynchronous communication with publish/subscribe) and \textit{TC2} (synchronous communication with client/server), two demo applications have been implemented. The application flow is realized via IEC~61499 applications, i.e., as FB networks. In order to increase the clarity of the diagrams, only the control logic is shown, the buttons are illustrated as inputs with the acronym \texttt{I\_XX} and the LEDs are represented as output with the acronym \texttt{Q\_XX}.

\vspace{2mm}

\subsubsection{Demand Side Management}

\begin{figure}[!htbp]
	\centering
		\includegraphics[width=0.98\columnwidth]{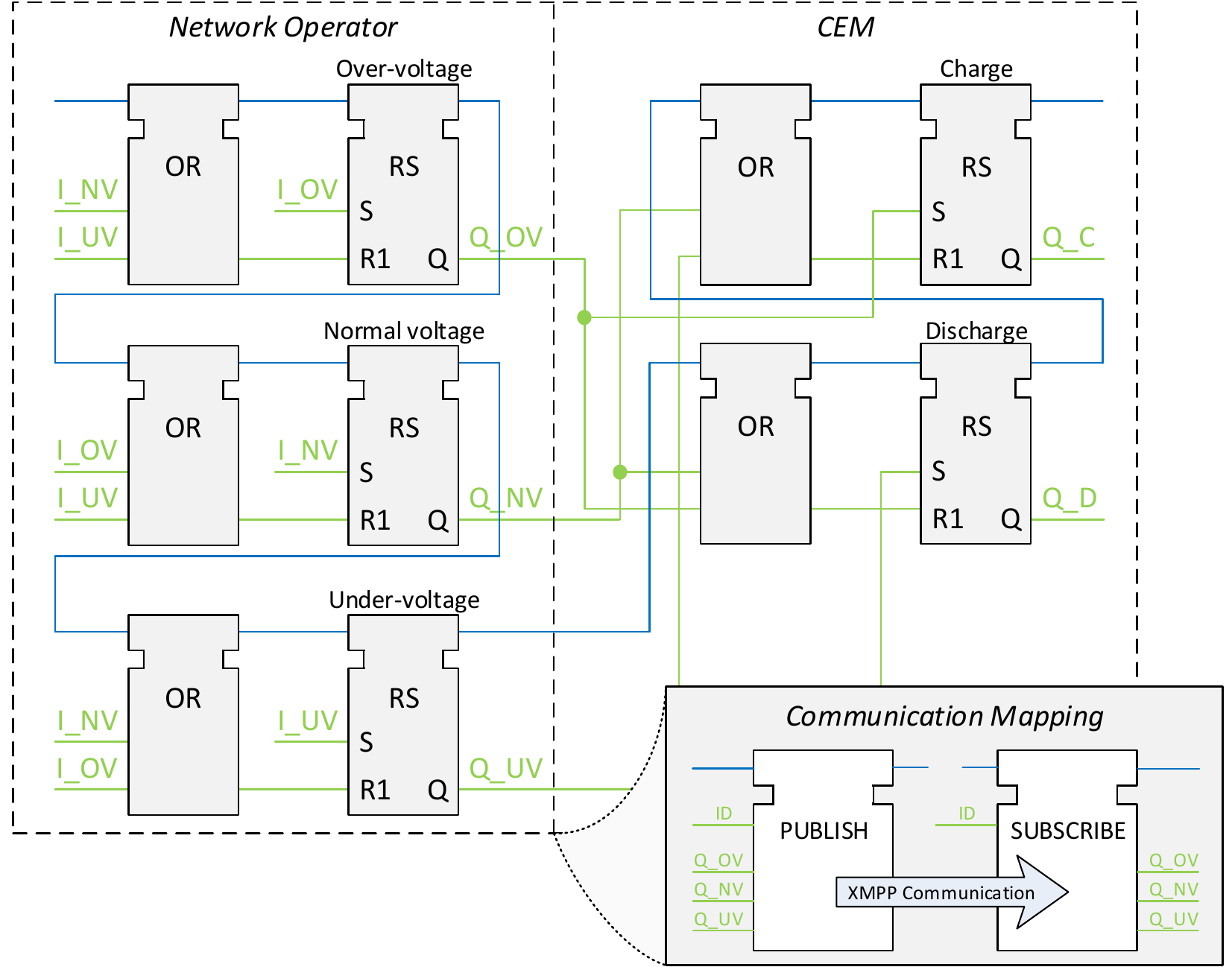}
	\caption{Demand Side Management application represented by IEC~61499 FBs (simplified view showing only the control logic).}
	\label{fig:tc1}
\end{figure}

The first IEC~61499 function block network is used for executing \textit{TC1} (see Fig.~\ref{fig:tc1}). Its left part is executed on the NetOp and the right part is executed on the CEM.

The left application part detects the voltage value of the grid via the three inputs (\texttt{I\_OV}, \texttt{I\_NV}, \texttt{I\_UV}). This state is internally saved via three RS flip-flops. The flip-flops are set when the corresponding button (e.g., \texttt{I\_NV}) is pressed and reset when one of the other two buttons (in this case \texttt{I\_OV} or \texttt{I\_UV}) is pressed. The two corresponding reset buttons are connected to the reset input of the flip-flop via an \texttt{OR} gate. The outputs are connected to the corresponding LEDs and the right application part.

The charge and discharge operations of the right application part are also realized via RS flip-flops. On over-voltage, the charge flip-flop is set and the discharge flip-flop is reset; on under-voltage the operation is vice versa. Both flip-flops are reset on normal voltage. The reset operation is also realized via \texttt{OR} gates. The outputs are connected to the corresponding LED for the charge (\texttt{Q\_C}) and discharge operation (\texttt{Q\_D}).

The reading of the three buttons is performed at an interval of 500 ms. On/off switching of the LEDs is performed immediately after setting the corresponding flip-flops. The transfer of the voltage values from NetOp to CEM is performed via a publish/subscribe (NetOp is the publisher, the CEM the subscriber) connection with XMPP Presence Stanzas.

\vspace{2mm}

\subsubsection{Reminder Function}

\begin{figure}[!htbp]
	\centering
		\includegraphics[width=0.98\columnwidth]{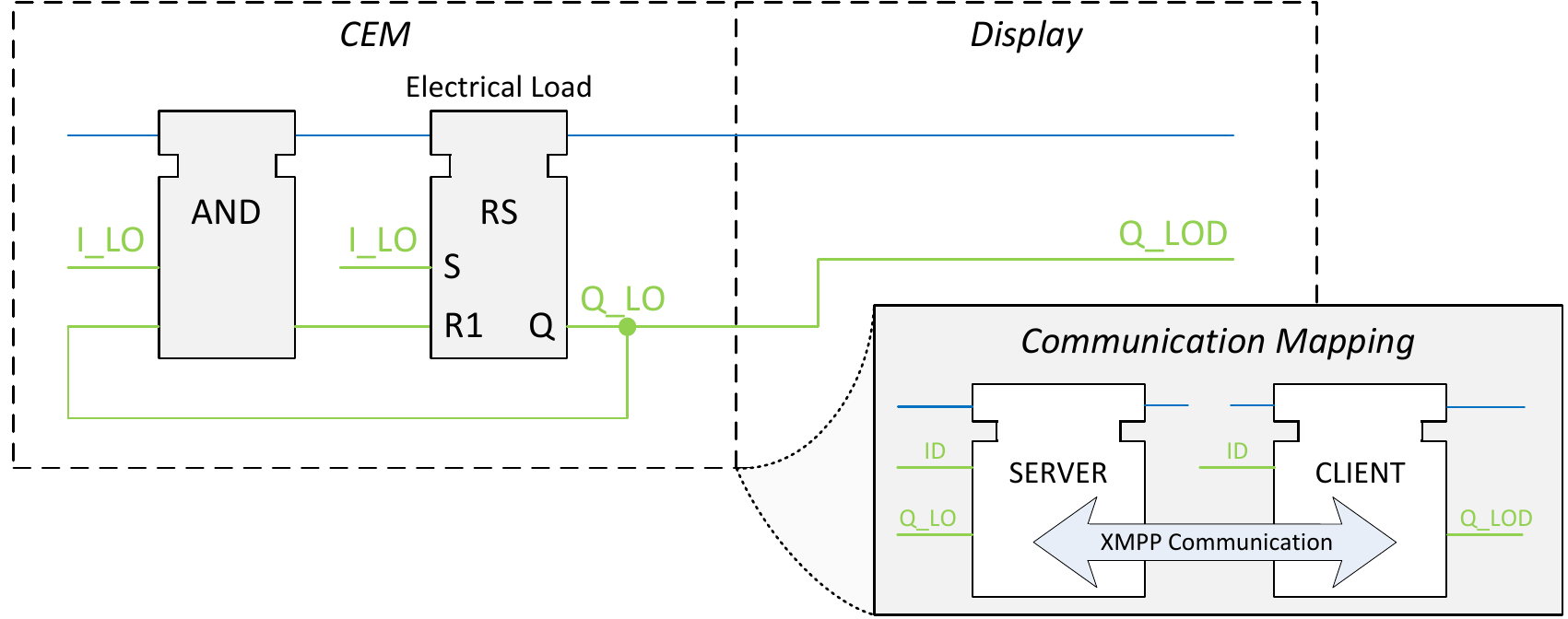}
	\caption{Reminder Function application represented by IEC~61499 FBs (simplified view showing only the control logic).}
	\label{fig:tc2}
\end{figure}

The second IEC~61499 application is used for the execution of \textit{TC2}, see Fig.~\ref{fig:tc2}. The left application part is executed on the CEM and the right part is executed on the display.

The left application part reads the state of the electrical load via the input \texttt{I\_LO}. This state is internally saved via an RS flip-flop. The flip-flop is set and reset when the button is pressed. The reset operation is realized via an \texttt{AND} gate, which connects the flip-flop output and the button with the flip-flop input. If the flip-flop is set and the button is pressed, the flip-flop will be reset. The output is connected to the power state LED \texttt{Q\_LO} and to the right application part. On the right application part, the power state is connected to the power state LED \texttt{Q\_LOD}.

The reading of the button is again performed at an interval of 500 ms. On/off switching of the LEDs is performed after setting the flip-flop. The transfer of the state represented by the flip-flop is performed via a client/server connection from the CEM to the display. The display requests the states with XMPP IQ Stanzas at an interval of one second, and the CEM responds accordingly. Thus, the display is the client and the CEM is the server here.

%The two application parts on the CEM (see Fig.~\ref{fig:componentdiagram}) are independently executed.

\subsection{Functional Validation}

For the functional validation of test case \textit{TC1}, the following main steps have been performed:

\begin{itemize}
	\item Set the current voltage value of the electricity grid (e.g., under-voltage) with the help of the three trigger buttons on the test board of the NetOp.
	\item Check the current state with the help of the three LEDs on the test board of the NetOp.
	\item Check the corresponding state of the buffer battery with the help of the two LEDs on the test board of the CEM.
\end{itemize}

In addition, the following steps have been performed for the functional validation of test case \textit{TC2}:

\begin{itemize}
	\item Set the power state of the electrical load with the help of the button on the test board of the CEM.
	\item Check the current power state with the help of the LED on the test board of the CEM.
	\item Check the power state with the help of the LED on the test board of the display.
\end{itemize}

As a result of the execution of the two test cases, it can be stated that all functional requirements have been completely fulfilled; thus, the prototype was validated against the functional requirements. Yet, further tests regarding the performance requirements (REQ 03 and 18) were necessary.

\subsection{Performance Evaluation}

The first performance test was an evaluation of the transmission time of the asynchronous communication (publish/ subscribe). The measurements were performed with the help of a digital oscilloscope. Channel one was connected to the over-voltage LED of the NetOp and channel two to the charging LED of the CEM. The measurements were executed for the communication methods XMPP encrypted, XMPP unencrypted, and UDP. The transmission time was 25 ms for XMPP encrypted, 21 ms for XMPP unencrypted, and 3 ms for UDP. The differences can be attributed to the overhead of XMPP and to the additional computing time the encryption and decryption needs.

The next performance test was a comparison of the payload size of the asynchronous and synchronous communication. The measurements were performed with the network sniffer Wireshark. The payload size is the sum of all messages that are required for the transmission of one value (for the client/server communication pattern: request, data transmission, acknowledgement). The communication methods XMPP encrypted, XMPP unencrypted, and UDP (for asynchronous communication) respectively TCP (for synchronous communication) were measured. The payload size for the asynchronous communication was 346 bytes for XMPP encrypted, 741 bytes for XMPP unencrypted and 45 bytes for UDP; for the synchronous communication, the payload size was 378 bytes for XMPP encrypted, 535 bytes for XMPP unencrypted and 202 bytes for TCP. The reason for the smaller payload of XMPP encrypted is that the XMPP library uses the zlib compression of OpenSSL.

The last performance test was a determination of the system load. The CPU load (in \% of the total usage) and the memory usage (in \% of the total usage and in kB of the Physical Memory (PM) and Virtual Memory (VM)) of FORTE after an execution time of 24 hours for each test have been analyzed (see Table~\ref{tab:system_load_XMPP_encrypted} to Table~\ref{tab:system_load_UDP_TCP}). Here, the three participating systems (NetOp, CEM, and display) have been evaluated when using XMPP encrypted, XMPP unencrypted and UDP respectively TCP.

\begin{table}[!htbp]
	\renewcommand{\arraystretch}{1.3}
	\caption{System load when using XMPP encrypted after 24 hours.}
	\label{tab:system_load_XMPP_encrypted}
	\centering
		\begin{tabular}{|l||c||c||c||c|}
			\hline
			\emph{System} & \emph{CPU} [\%] & \emph{Mem} [\%] & \emph{PM} [kB] & \emph{VM} [kB]\\
			\hline
			NetOp & 6.0 & 10.9 & 48,872 & 92,624\\
			\hline
			CEM & 7.9 & 11.1 & 49,780 & 110,584\\
			\hline
			Display & 4.0 & 1.5 & 6,912 & 50,040\\
			\hline
		\end{tabular}
\end{table}

\begin{table}[!htbp]
	\renewcommand{\arraystretch}{1.3}
	\caption{System load when using XMPP unencrypted after 24 hours.}
	\label{tab:system_load_XMPP_unencrypted}
	\centering
		\begin{tabular}{|l||c||c||c||c|}
			\hline
			\emph{System} & \emph{CPU} [\%] & \emph{Mem} [\%] & \emph{PM} [kB] & \emph{VM} [kB]\\
			\hline
			NetOp & 5.7 & 10.5 & 46,992 & 91,488\\
			\hline
			CEM & 8.3 & 10.5 & 47,128 & 108,792\\
			\hline
			Display & 4.6 & 1.1 & 4,980 & 49,292\\
			\hline
		\end{tabular}
\end{table}

\begin{table}[!htbp]
	\renewcommand{\arraystretch}{1.3}
	\caption{System load when using UDP/TCP after 24 hours.}
	\label{tab:system_load_UDP_TCP}
	\centering
		\begin{tabular}{|l||c||c||c||c|}
			\hline
			\emph{System} & \emph{CPU} [\%] & \emph{Mem} [\%] & \emph{PM} [kB] & \emph{VM} [kB]\\
			\hline
			NetOp & 4.8 & 0.9 & 4,084 & 40,784\\
			\hline
			CEM & 6.8 & 0.9 & 4,328 & 48,944\\
			\hline
			Display & 4.0 & 0.9 & 4,072 & 40,704\\
			\hline
		\end{tabular}
\end{table}

The CPU load with asynchronous communication after a period of 24 hours was about one percent higher for XMPP encrypted and unencrypted than for UDP; with synchronous communication the CPU load was as high as for TCP. The memory usage with asynchronous communication was about ten times higher for XMPP encrypted and XMPP unencrypted than for UDP; with synchronous communication the memory usage was about 50\% higher for XMPP encrypted and unencrypted than for TCP. 

The performance test was continued over a period of three days (72 hours). The result was that the memory usage of the publish/subscribe communication for XMPP encrypted (see Table \ref{tab:system_load_XMPP_encrypted_three_days}) and XMPP unencrypted was further increasing. The reason for this behavior is possibly a memory leak in FORTE or in the XMPP library.

\begin{table}[!htbp]
	\renewcommand{\arraystretch}{1.3}
	\caption{System load when using XMPP encrypted after 72 hours}
	\label{tab:system_load_XMPP_encrypted_three_days}
	\centering
		\begin{tabular}{|l||c||c||c||c|}
			\hline
			\emph{System} & \emph{CPU} [\%] & \emph{Mem} [\%] & \emph{PM} [kB] & \emph{VM} [kB]\\
			\hline
			NetOp & 5.8 & 30.8 & 137,588 & 180,816\\
			\hline
			CEM & 6.4 & 30.9 & 138,008 & 198,176\\
			\hline
			Display & 4.8 & 1.6 & 7,520 & 50,720\\
			\hline
		\end{tabular}
\end{table}

Summarizing, the performance tests show that the transmission time of XMPP is about eight times higher than UDP, the payload size is about 17 times larger than UDP and about three times larger than TCP, the CPU load is about one percent higher and the memory usage is about ten times higher than both UDP and TCP.

The measured performance disadvantages are a drawback of XMPP based solutions and should be reduced. Possible improvements could be the usage of XMPP over UDP and XMPP message compression (e.g., binary XML). On the other hand, XMPP also offers many advantages and useful features:

\begin{itemize}
	\item{The XMPP protocol is expandable via XMPP extensions.}
	\item{XMPP messages use XML for the message format. Therefore XMPP messages can contain hierarchical data structures and semantic models.}
	\item{XMPP messages can be exchanged between heterogeneous systems because XMPP messages are text-based (Unicode) messages in XML format.}
	\item{XMPP supports asynchronous, event-based and synchronous communication.}
	\item{The XMPP protocol supports encryption via TLS and authentication.}
	\item{XMPP servers administrate the user accounts, user rights and security policies.}
	\item{XMPP can address local (LAN) and global nodes (WAN). Nodes are addressed via Jabber Identifier (JID).}
	\item{The XMPP communication network can be divided into several parts. Each part is controlled by an individual XMPP server.} 
	\item{New XMPP nodes can be added very easily. The new node only requires the JID of the communication partner. An automatic service discovery system could be implemented with the XMPP extension ``XEP-0030: Service Discovery''.}
\end{itemize}

Ultimately, it has to be decided on a case-by-case basis, whether or not the functional advantages outweigh the performance disadvantages.

%% file: Conclusion.tex
\section{Conclusions and Further Work}
\label{sec:conclusions}

% The conclusion summarizes all before mentioned chapters and gives some remarks about the degree of success compared with the overall goals defined in the introduction.

% Furthermore some outlook to potential further research activities and/or planned dissemination and exploitation plans.

In this paper, an XMPP-based communication infrastructure for distributed smart grid applications has been proposed. Requirements and defined test cases have been collected in order to validate the proposed solution against the fulfillment of these requirements. Subsequently, the architecture of a prototypical solution has been developed and the  prototype has been implemented accordingly. In the evaluation part, not only the validity of the used approach could be proven, but also a first qualitative evaluation of the solution at hand could be given. As a result, it can be seen that the feasibility of the concept and the elegance of the solution for interoperability contrast with some drawbacks in performance.

Thus, future research and development activities have to prove, whether or not XMPP based solutions are viable for large scale roll out of distributed smart grid applications. Some performance improvements may be expected by using XMPP message compression. Also, alternative solutions such as MQTT or CoAP should be investigated. Furthermore, the work on engineering technologies for designing and implementing IEC~61499 based applications has to be intensified in order to prove large scale feasibility. For this part, usability issues should be taken into account as well.